\documentclass[12pt,american]{article}
\usepackage[T1]{fontenc}
\usepackage{textcomp}
\usepackage[latin9]{inputenc}
\usepackage{xcolor}
\usepackage{babel}
\usepackage{amsmath}
\usepackage{amssymb}
\usepackage{graphicx}
\usepackage{geometry}
\usepackage{wasysym}
\usepackage[pdfusetitle,
 bookmarks=true,bookmarksnumbered=false,bookmarksopen=false,
 breaklinks=false,pdfborder={0 0 1},backref=false,colorlinks=false]
 {hyperref}

\makeatletter
\@ifundefined{date}{}{\date{}}
\usepackage{soul}

\makeatother

\begin{document}
\title{The continuum limit of the Poland-Scheraga DNA denaturation model}
\author{R. Dengler \thanks{ORCID: 0000-0001-6706-8550}\\
Munich, Germany}
\maketitle
\begin{abstract}
Using a field theory equivalent to a lattice version of the Poland-Scheraga
model, the phase diagram for a long DNA molecule is derived in closed
form.

For the generalized model with excluded-volume interactions a one-loop
renormalization group calculation shows that there are two stable
fixed points. At both fixed points, the excluded-volume effect plays
a role. At the fixed point reached when the original excluded-volume
effect is weak, the phase transition is continuous. At the other fixed
point, the phase transition is first order.
\end{abstract}

\section{Introduction}

\begin{figure}
\centering
\includegraphics[scale=0.6]{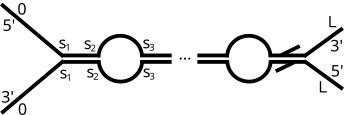}\caption{\label{fig:Scheraga_Gamma}In the PS model two complementary DNA single
strands can pair only at same distances $s$ from the endpoints. The
length is measured from $3'$ to $5'$ on the template strand and
from $5'$ to $3'$ on the other strand. The only allowed conformations
are a sequence of loops.}
\end{figure}

The subject of the Poland-Scheraga (PS) \cite{Poland_1966} model
is a long DNA molecule in a good solvent. When the temperature is
increased, the DNA molecule denatures. It is assumed that the pairing
energy does not depend on the type of nucleotides and that nucleotides
of two complementary DNA single strands can pair only when their length
indices along the two strands agree. This suggests measuring the length
of the template strand from 3\textasciiacute{} to 5\textasciiacute{}
and the length of the other strand from 5\textasciiacute{} to 3\textasciiacute{}
(see Fig.~\ref{fig:Scheraga_Gamma}). The two DNA single strands
can then form a sequence of double strands and loops.

The model captures essential biological and physical aspects of denaturation
in a semi-quantitative way. In the original PS model, excluded-volume
effects are ignored. The model and its generalized (gPS) versions
are the subject of new studies, often with a mathematical background.
An introduction can be found in~\cite{Richard_2004}. Generalized
versions take excluded-volume effects into account~\cite{Pelity_2000,Garel_2001},
allow a length-dependent pairing energy \cite{Berger_2024}, or permit
some pairing at not exactly complementary length indices~\cite{Garel_2004,Giaco_2017}. 

The purpose of this work is to examine the continuum limit of the
PS model and of a gPS model with excluded-volume effect. The term
\textquotedbl continuum limit\textquotedbl{} here means that the
polymers are long and that stiffness effects can be ignored. The correlation
length of long, dilute linear polymers in a good solvent usually diverges
as $\xi\sim\ell^{\nu}$ with polymer length $\ell$, where $\nu$
is a universal exponent~\cite{Gennes1972,Lubensky1979,Dengler2024}.
The large correlation length is a hallmark of critical phenomena,
and it is of interest to identify the universality classes and the
universal properties of such polymer solutions.

The central tool used in this work is the field theory corresponding
to the gPS model in the continuum limit, which can be derived in a
formal way from a lattice model \cite{Dengler2025}. We do not repeated
the derivation here. The crucial point for the interpretation of the
field theory is the connection between its harmonic part and Gaussian
curves of given length. This connection is derived at the level of
a lattice model in appendix \ref{subsec:FT_Lattice}. In a perturbation
series the nonlinearities then connect Gaussian curves in the usual
way, and Feynman diagrams and polymer conformations have exactly the
same topology.

The action integral corresponding to the PS model has the form
\begin{align}
S & =\int_{x,s}\tilde{\varphi}\left[r_{0}-\nabla^{2}+\partial_{s}\right]\varphi+\int_{x,s}\tilde{\psi}\left[\tau_{0}-\nabla^{2}+W\partial_{s}\right]\psi-\tfrac{\lambda}{\sqrt{K_{d}}}\int_{x,s}\left(\psi\tilde{\varphi}^{2}+\varphi^{2}\tilde{\psi}\right)\label{eq:Act}\\
 & \qquad+\tfrac{u_{\varphi}}{K_{d}}\int_{x}\left(\int_{s}\tilde{\varphi}\varphi\right)^{2}+\tfrac{u_{\psi}}{K_{d}}\int_{x}\left(\int_{s}\tilde{\psi}\psi\right)^{2}+2\tfrac{u_{\varphi\psi}}{K_{d}}\int_{x}\left(\int_{s}\tilde{\varphi}\varphi\right)\left(\int_{s}\tilde{\psi}\psi\right).\nonumber 
\end{align}
The meaning of the symbols is as follows. The variable $s$ measures
the length along the polymer strands. The integrals are abbreviations
for $\int_{x,s}\ldots=\int\mathrm{d}^{d}x\mathrm{d}s\ldots$, the
factors $K_{d}=2^{1-d}\pi^{-d/2}/\Gamma\left(d/2\right)$ eliminate
complicated factors later on~\cite{Amit1978}. 

A double strand field $\psi\left(\boldsymbol{x},s\right)$ usually
depends on two length variables, but one of them is redundant in the
PS model and has been omitted. The single strand field $\varphi\left(\boldsymbol{x},s\right)$
has one length variable. The directed length $s$ along polymer strands
requires pairs of fields, polymer sources ($\tilde{\varphi},\tilde{\psi}$)
and polymer sinks ($\varphi,\psi$).

The nonlinear parts with coupling constant $\lambda$ describe denaturation
of a double strand $\psi$ to two single strands $\varphi$ and the
reverse process. In the lattice version of the model, $\lambda$ is
the weight with which conformations in which three endpoints coincide
actually combine to a vertex. It is a peculiarity of the PS model
that all interactions take place at identical length variables. The
first line of (\ref{eq:Act}) accordingly is local in $s$, and the
length variable could also be interpreted as time. The action (\ref{eq:Act})
with $u=0$ resembles that of directed percolation, and formally also
is a model of critical dynamics~\cite{HH77}. One might attempt to
interpret the PS model as a reaction-diffusion process of two types
of diffusing particles, with reaction formulas $P\rightarrow2F$ and
$2F\rightarrow P$. However, such a process generates additional marginal
and relevant terms~\cite{Pelity_1985}.

The nonlinear parts with coupling constant $u$ describe the excluded-volume
effect, are nonlocal in $s$ and not part of the standard PS model.
The length integrals mean that two strands interact in the same way
for any combination of length variables.

If one attributes a Boltzmann factor $e^{\beta E_{\varphi}}$ to each
$\varphi$ monomer and $e^{\beta E_{\psi}}$ to each $\psi$ monomer
then the lattice model generates the parameters
\begin{align}
r_{0} & =c_{\varphi}-\beta E_{\varphi},\label{eq:r0_def}\\
\tau_{0} & =c_{\psi}-\beta E_{\psi},\label{eq:tau0_def}
\end{align}
where $c_{i}$ are non-universal constants, see also appendix \ref{subsec:FT_Lattice}.

A positive energy $E_{j}$ favors the respective polymer type, and
the parameters $r_{0}$ and $\tau_{0}$ thus control the degree of
denaturation. Since $m_{\varphi}=\int_{x,s}\tilde{\varphi}\varphi$
is the $\varphi$-mass the weight $e^{-S}$ in a natural way contains
the factor $e^{\beta E_{\varphi}m_{\varphi}}$, and likewise for $m_{\psi}$
and $E_{\psi}$.

Instead of starting with the action (\ref{eq:Act}), one could use
a field theory of the type introduced by Edwards, based on polymer
paths, and manually add length-dependent Boltzmann factors and interactions.
Alternatively, one could simply combine Gaussian functions for the
noninteracting polymer segments. The advantage of an action like (\ref{eq:Act})
is that it follows from a lattice model and makes available the field
theoretic tools and concepts.

\section{Physical quantities}

The correlation function $G\left(\boldsymbol{x},s\right)=\left\langle \tilde{\psi}\left(0,0\right)\psi\left(\boldsymbol{x},s\right)\right\rangle $
is the statistical weight generated by the action (\ref{eq:Act})
for a double strand polymer starting at the origin with length $0$
and terminating at position $\boldsymbol{x}$ with length $s$. The
Fourier transform $G\left(\boldsymbol{k}=0,s\right)=\int_{x}\left\langle \tilde{\psi}\left(0,0\right)\psi\left(\boldsymbol{x},s\right)\right\rangle $
is the statistical weight generated by the action (\ref{eq:Act})
for a double strand polymer of length $s$ starting at the origin
and terminating anywhere. The average single strand mass $m_{\varphi}$
in the latter case is
\[
\left\langle m_{\varphi}\right\rangle =-\partial G\left(\boldsymbol{k}=0,s\right)/\partial r_{0},
\]
and analogously for $m_{\psi}$ and $\tau_{0}$. This, however, are
not yet physical quantities.  The correlation function $G\left(\boldsymbol{x},s\right)$
in its lattice model version is the sum over all conformations $C\left(\boldsymbol{x},s\right)$
with length $s$ and given endpoints of Boltzmann factors for the
polymer masses $m_{\varphi}$ and $m_{\psi}$, 
\[
G\left(\boldsymbol{x},s\right)=\sum_{\alpha\in C\left(\boldsymbol{x},s\right)}e^{-m^{(\alpha)}_{\varphi}r_{0}-m^{(\alpha)}_{\psi}\tau_{0}}=e^{-m\tau_{0}}\sum_{\alpha\in C\left(\boldsymbol{x},s\right)}e^{-m^{(\alpha)}_{\varphi}\left(r_{0}-\tau_{0}\right)}.
\]
Here it was used that the given total length $s$ enforces a constant
total mass $m=m^{(\alpha)}_{\varphi}+m^{(\alpha)}_{\psi}$. The Boltzmann
factors in the sum are the correct weight for the conformations according
to some pairing energy.  The unphysical overall factor $e^{-m\tau_{0}}$
drops out from the normalized probability distribution
\[
P\left(\boldsymbol{x},s\right)=G\left(\boldsymbol{x},s\right)/\sum_{\boldsymbol{x}}G\left(\boldsymbol{x},s\right).
\]
The properly normalized average single strand mass thus is
\begin{equation}
\overline{m_{\varphi}}\left(s,r_{0},\tau_{0}\right)=-\partial\ln G\left(\boldsymbol{k}=0,s\right)/\partial r_{0}.\label{eq:m_phi_normalized}
\end{equation}

\section{Loop exponent}

\label{sec:PS_Loop_Exp_Def}To derive the loop exponent $c$, a central
quantity in the discrete PS model, we write the $\varphi$-propagator
$g_{0}\left(\boldsymbol{x},s\right)=\left\langle \tilde{\varphi}\left(\boldsymbol{0},0\right)\varphi\left(\boldsymbol{x},s\right)\right\rangle _{0}$
in two forms
\begin{align}
g_{0}\left(\boldsymbol{x},s\right) & =2^{-d}\pi^{-d/2}\theta\left(s\right)s^{-d/2}e^{-r_{0}s-x^{2}/4s},\label{eq:g0}\\
g_{0}\left(\boldsymbol{k},\omega\right) & =1/\left(r_{0}+k^{2}-i\omega\right).\label{eq:g0_omega}
\end{align}
The first line is a normalized diffusion curve with an additional
Boltzmann factor $e^{-r_{0}s}.$ It measures the probability to find
the end of a single strand polymer of length $s$ at $\boldsymbol{x}$
when the other end is fixed at the origin. The second line is the
Fourier transform $g\left(\boldsymbol{k},\omega\right)=\int_{x,s}e^{-i\boldsymbol{kx}+i\omega s}g_{0}\left(\boldsymbol{x},s\right)$,
where the ``frequency'' $\omega$ is the variable conjugate to $s$. 

The statistical weight $\Omega\left(s\right)$ of a $\varphi$-loop
of length $s$ of the type shown in Fig.~\ref{fig:Scheraga_Gamma}
is given by $\Omega\left(s\right)=\int_{x}g^{2}_{0}\left(\boldsymbol{x},s\right)\propto e^{-2r_{0}s}s^{-d/2}$.
This ``loop weight'' plays a central role in the discrete formulation
of the PS model, and for large $s$ it is assumed $\Omega\left(s\right)\sim K^{s}s^{-c}$
with some constant $K$ and a ``loop exponent'' $c$. For the PS
model variant with $u=0$ it follows $c=d/2$. According to a central
PS theorem valid also for more generic variants of the PS model, the
denaturation is a continuous phase transition for $1\leq c<2$.

\section{Exact solution without excluded-volume effects}

The single strand propagator $\left\langle \tilde{\varphi}\varphi\right\rangle $
is exactly given by (\ref{eq:g0}). The double strand propagator is
determined by the PS loop, the one-loop vertex function or self energy
\begin{align}
\Gamma^{(1)}_{\tilde{\psi}\psi}\left(\boldsymbol{k},\omega\right) & =-\tfrac{2\lambda^{2}}{K_{d}\left(2\pi\right)^{d}}\int^{\Lambda}\tfrac{\mathrm{d}^{d}p}{p^{2}+\left(k-p\right)^{2}+2r_{0}-i\omega}\label{eq:Gamm2}\\
 & =\lambda^{2}\left(\tfrac{\pi/2}{\sin\pi\epsilon/2}\left(r_{0}+\tfrac{k^{2}}{4}-\tfrac{i\omega}{2}\right)^{1-\epsilon/2}-\tfrac{\Lambda^{2-\epsilon}}{2-\epsilon}-\tfrac{\Lambda^{-\epsilon}}{\epsilon}\left(r_{0}+\tfrac{k^{2}}{4}-\tfrac{i\omega}{2}\right)+\ldots\right).\nonumber 
\end{align}
In the first line the internal $\omega$-integral has already been
performed, and $\Lambda=a^{-1}$ is the UV cutoff, where $a$ is the
lattice spacing of the original lattice model. As usual $\epsilon=4-d$. 

The vertex function (\ref{eq:Gamm2}) is a sum of a singular contribution,
a constant, and non-universal contributions regular in the parameters
$r_{0}$, $k$ and $\omega.$ The constant is uninteresting, it only
shifts the parameter $\tau_{0}$. The regular contributions of higher
order contain additional powers of $r_{0}/\Lambda^{2}$, $k^{2}/\Lambda^{2}$
and $i\omega/\Lambda^{2}$ and thus are negligible near the critical
point (for lengths much larger than the lattice constant $1/\Lambda$).
The double strand propagator $\left\langle \tilde{\psi}\psi\right\rangle $
follows as
\begin{align}
G\left(k,\omega\right) & =\left(\tau_{0}+k^{2}-Wi\omega+\Gamma^{(1)}_{\tilde{\psi}\psi}\right)^{-1}\label{eq:propPsi}\\
 & \cong\left[\tau_{0}+k^{2}-Wi\omega-\lambda^{2}\tfrac{\Lambda^{-\epsilon}}{\epsilon}\left(r_{0}+\tfrac{k^{2}}{4}-\tfrac{i\omega}{2}\right)+\lambda^{2}\tfrac{\pi/2}{\sin\pi\epsilon/2}\left(r_{0}+\tfrac{k^{2}}{4}-\tfrac{i\omega}{2}\right)^{1-\epsilon/2}\right]^{-1}.\nonumber 
\end{align}
The geometric series in the coupling constant $\lambda^{2}$ is typical
for the PS model. For long polymers the typical values of $\tau_{0}$,
$r_{0},$ $k^{2}$ and $\omega$ are small, and the singular term
is larger by a factor $k^{-\epsilon}\sim\xi^{\epsilon}$. 

\subsection{Crossover}

\begin{figure}
\centering
\includegraphics[scale=0.5]{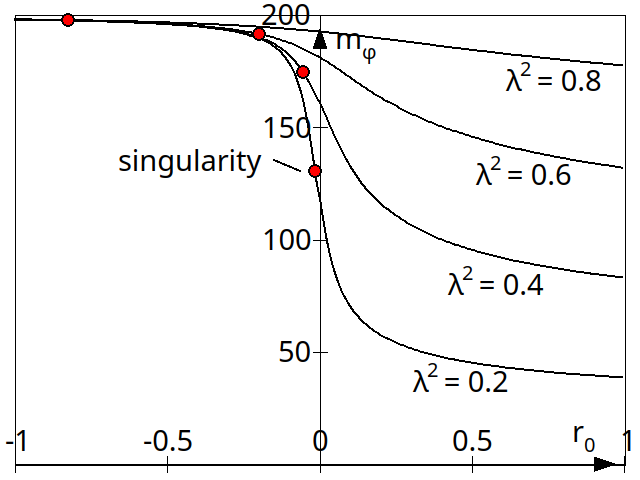}\caption{\label{fig:m_phi}Single strand mass $m_{\varphi}$ as a function
of pairing energy $r_{0}=c_{\varphi}-\beta E_{\varphi}$ for fixed
total length 200, $W=1$ and various coupling constant values $\lambda^{2}$,
calculated with dimensional regularization. }
\end{figure}
We now set $\boldsymbol{k}=0$, which means that the double strand
endpoint can be anywhere. Physical properties can be derived after
a Fourier transformation of (\ref{eq:propPsi}) from $\omega$ space
to length space $s$. For the most interesting dimension $d=3$ this
can be done in closed form,
\begin{align}
G_{3\mathrm{d}}\left(\boldsymbol{k}=0,s\right) & =\mathrm{const}\times\int^{\infty}_{-\infty}\tfrac{\mathrm{d}\omega}{2\pi}e^{-i\omega s}\frac{1}{T\left(r_{0},\tau_{0}\right)-i\omega+L\sqrt{2r_{0}-i\omega}}\label{eq:G_3d_s}\\
 & \overset{d=3}{\propto}\frac{e^{-2r_{0}s}}{y_{1}-y_{2}}\left(y_{1}e^{y^{2}_{1}s}\mathrm{erfc}\left(-y_{1}\sqrt{s}\right)-y_{2}e^{y^{2}_{2}s}\mathrm{erfc}\left(-y_{2}\sqrt{s}\right)\right),\nonumber \\
y_{1,2} & =\tfrac{1}{2}\left(-L\pm\sqrt{L^{2}+8r_{0}-4T\left(r_{0},\tau_{0}\right)}\right).\nonumber 
\end{align}
A derivation of the integral can be found in appendix~\ref{subsec:FT_integral}.
The function $\mathrm{erfc}=1-\mathrm{erf}$ is the complementary
error function. One could rescale the integration variable in (\ref{eq:G_3d_s})
like $\omega\rightarrow\Omega/s$. This confirms that higher powers
of $\omega$ in the denominator are negligible when the length $s$
is large. The overall constant factor and the constant $L\left(\lambda^{2},W,\epsilon\right)$
only define non-universal scales and are uninteresting. The parameter
$T\left(\lambda^{2},W,\epsilon,\tau_{0},r_{0}\right)$ is linear in
$\tau_{0}$ and $r_{0}$ near the critical point. For $\left|r_{0}\right|\cong\Lambda^{2}$
higher terms in $r_{0}/\Lambda^{2}$ are not negligible any more in
(\ref{eq:Gamm2}) and one must include such terms or use dimensional
regularization.

\subsubsection{Single strand mass}

The average normalized single strand mass (\ref{eq:m_phi_normalized})
follows as
\[
\overline{m_{\varphi}}=-\partial\ln G_{3\mathrm{d}}\left(0,s\right)/\partial r_{0}.
\]
The derivative can be determined numerically. One needs an implementation
of the $\mathrm{erfc}$ function for complex arguments. The $\varphi$
mass is plotted as a function of $r_{0}$ from Eq.(\ref{eq:r0_def})
in Fig.~\ref{fig:m_phi}. At the critical point of the model a finite
fraction of the DNA molecule is denatured. The exact solution (\ref{eq:G_3d_s})
for $d=3$ also describes the crossover. 

\section{Renormalization group calculation}

For $r_{0}=\tau_{0}=0$ a dimensional analysis shows that the action
(\ref{eq:Act}) formally is scale invariant with an upper critical
dimension $d_{c}=4$, which naturally calls for a renormalization
group (RG) treatment. The required techniques are standard~\cite{Amit1978}.
However, because the algebra with four field types and four coupling
constants can become involved, we summarize all essential details.

\subsection{Standard PS model}

We now use the exact solution (\ref{eq:Gamm2}) for $u=0$ to introduce
the renormalization group (RG) formalism. This does not lead to new
results, but it is a good preparation for the more complicated case
with $u\neq0.$

The usual arguments are as follows. The action integral (\ref{eq:Act})
in combination with different cutoffs $\Lambda$ represents instances
of the same universality class, with the same critical exponents.
The \emph{amplitudes} of vertex functions, however - such as $\Gamma_{\tilde{\varphi}\varphi}=r_{0}+k^{2}-i\omega$
and $\Gamma_{\tilde{\psi}\psi}=\tau_{0}+k^{2}-i\omega+\Gamma^{(1)}_{\tilde{\psi}\psi}$,
are non-universal. To eliminate the non-universal aspects one rescales
the fields and coordinates according to $\varphi=Z_{\varphi}\varphi_{\mathrm{R}},$
$\tilde{\varphi}=Z_{\varphi}\tilde{\varphi}_{\mathrm{R}},$ $\psi=Z_{\psi}\psi,$
$\tilde{\psi}=Z_{\psi}\tilde{\psi}_{\mathrm{R}}$ and $s=Z_{s}s_{\mathrm{R}}$
and imposes normalization conditions at some arbitrary small wavevector
$\mu$, for instance 
\begin{align}
\partial\Gamma^{\mathrm{R}}_{\tilde{\varphi}\varphi}/\partial k^{2}|_{k=\mu,\omega=0} & =1,\label{eq:RenCondPhi2k}\\
\partial\Gamma^{\mathrm{R}}_{\tilde{\varphi}\varphi}/\partial\left(-i\omega\right)|_{k=\mu,\omega=0} & =1,\label{eq:RenCondPhi2Omega}\\
\partial\Gamma^{\mathrm{R}}_{\tilde{\psi}\psi}/\partial k^{2}|_{k=2\mu,\omega=0} & =1.\label{eq:RenCondPsi2k}
\end{align}
The symbol $\mathrm{R}$ denotes renormalized quantities. The renormalized
vertex functions get their $Z$-factors from the attached fields and
thus (using dimensional regularization) 
\begin{align*}
\Gamma^{\mathrm{R}}_{\tilde{\varphi}\varphi}\left(k,\omega\right) & =Z^{2}_{\varphi}Z_{s}\left[k^{2}-Z^{-1}_{s}i\omega\right],\\
\Gamma^{\mathrm{R}}_{\tilde{\psi}\psi}\left(k,\omega\right) & =Z^{2}_{\psi}Z_{s}\left[k^{2}-Z^{-1}_{s}i\omega+\lambda^{2}\tfrac{\pi/2}{\sin\pi\epsilon/2}\left(\tfrac{k^{2}}{4}-Z^{-1}_{s}\tfrac{i\omega}{2}\right)^{1-\epsilon/2}\right].
\end{align*}
The renormalization conditions lead to
\begin{align*}
1 & =Z_{\varphi}=Z_{s},\\
1 & =Z^{2}_{\psi}Z_{s}\left[1+\tfrac{\bar{\lambda}^{2}}{4}\tfrac{\pi/2}{\sin\pi\epsilon/2}\left(1-\tfrac{\epsilon}{2}\right)\right],
\end{align*}
where $\bar{\lambda}=\lambda\mu^{-\epsilon/2}$ is the dimensionless
bare coupling constant. With the given $Z$ factors the dimensionless
renormalized coupling constant $\lambda_{R}$ follows as
\[
\lambda_{R}\left(\bar{\lambda}\right)\equiv\mu^{-\epsilon/2}\Gamma^{R}_{\varphi\varphi\tilde{\psi}}\left(\mu,\lambda\right)=Z^{2}_{\varphi}Z_{\psi}Z_{s}\bar{\lambda}.
\]
Inserting the factors yields
\begin{equation}
\lambda^{2}_{R}\left(\bar{\lambda}\right)=Z^{2}_{\psi}\bar{\lambda}^{2}=\tfrac{\bar{\lambda}^{2}}{1+\tfrac{\bar{\lambda}^{2}}{4}\tfrac{\pi/2}{\sin\pi\epsilon/2}\left(1-\tfrac{\epsilon}{2}\right)}\overset{\mu\rightarrow0}{\longrightarrow}4\tfrac{\sin\pi\epsilon/2}{\left(\pi/2\right)\left(1-\epsilon/2\right)}.\label{eq:lambda_R_exact}
\end{equation}
We thus have found a stable fixed point $\lambda^{2}_{*\mathrm{R}}=4\epsilon+O\left(\epsilon^{2}\right).$
This critical point is reached when $\varphi$ and $\psi$ are ``massless'',
that is if both types of polymer strands are long. The $\lambda^{2}_{\mathrm{R}}$
remains finite also for $d=2$.

The conditions (\ref{eq:RenCondPhi2k}, \ref{eq:RenCondPhi2Omega},
\ref{eq:RenCondPsi2k}) are sufficient to determine the scale factors
$Z$ and $\lambda_{R}$. One can now also calculate 
\[
W_{\mathrm{R}}=\partial\Gamma^{\mathrm{R}}_{\tilde{\psi}\psi}\left(k,\omega\right)/\partial\left(-i\omega\right)\overset{\mu\rightarrow0}{\longrightarrow}2\mu^{\epsilon}\left(\tfrac{k^{2}}{4}-\tfrac{i\omega}{2}\right)^{-\epsilon/2}.
\]
This quantity flows to the fixed point $W_{\mathrm{R}}=2$ at the
normalization points $k=2\mu$ and $\omega=0$ or $k=0$ and $\omega=i\mu^{2}/2$.

\subsection{Excluded-volume interaction}

\begin{figure}
\centering
\includegraphics[scale=0.7]{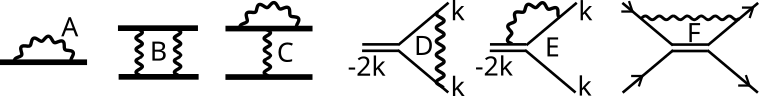}\caption{\label{fig:OneLoopExcl}One loop renormalizations due to the excluded
-volume interactions (wiggly line). There are no \textquotedblleft frequency\textquotedblright{}
integrals. The thick lines in the diagrams A, B and C can be a single
strand or a double strand. The wavevector $k$ indicates the symmetry
point used in the renormalization. The $\psi$-line in diagram F constricts
the length variables of $\varphi$, leading to an irrelevant contribution.}
\end{figure}
The excluded-volume interactions with coupling constants $u$ generate
the one-loop diagrams shown in Fig.~\ref{fig:OneLoopExcl}. The single
strand sector of (\ref{eq:Act}) still is not affected by the double
strand $\psi$ and one finds
\begin{align}
Z_{\varphi} & =1+\tfrac{1}{\epsilon}\bar{u}_{\varphi},\qquad Z_{s}=1-\tfrac{2}{\epsilon}\bar{u}_{\varphi},\label{eq:Z_phi_Z_s}\\
\beta\left(u^{\mathrm{R}}_{\varphi}\right) & =\mu\partial_{\mu}u^{\mathrm{R}}_{\varphi}=u^{\mathrm{R}}_{\varphi}\left(-\epsilon+8u^{\mathrm{R}}_{\varphi}\right).\label{eq:beta_u_phi}
\end{align}
Here $\bar{u}_{\varphi}=u_{\varphi}\mu^{-\epsilon}$ and $u^{R}_{\varphi}$
denote the dimensionless bare and renormalized coupling constant.
The fixed point is $u^{R}_{\varphi}=\epsilon/8$. The usual critical
exponent $\nu$ follows from 
\begin{equation}
\mu\partial_{\mu}\ln Z_{s}\cong2u^{\mathrm{R}}_{\varphi}=2-1/\nu\label{eq:ExpNu}
\end{equation}
as $\nu\cong\tfrac{1}{2}+\tfrac{\epsilon}{16}$, in agreement with
de Gennes~\cite{Gennes1972}. 

It remains to examine the mixed sector. The diagram A contributes
to $\Gamma^{R}_{\tilde{\psi}\psi}$. However, it does not depend on
$k^{2}$ and thus in leading order in $\epsilon$ 
\begin{align}
1 & =Z^{2}_{\psi}Z_{s}\left[1+\tfrac{\bar{\lambda}^{2}}{4\epsilon}\right],\qquad Z_{\psi}=1-\tfrac{1}{8\epsilon}\bar{\lambda}^{2}+\tfrac{1}{\epsilon}\bar{u}_{\varphi}.\label{eq:Z_Psi}\\
\eta_{\psi} & \equiv\mu\partial_{\mu}\ln Z_{\psi}=\tfrac{1}{8}\bar{\lambda}^{2}-\bar{u}_{\varphi}.\label{eq:ExpEtaPsi}
\end{align}
The critical exponent $\eta_{\psi}$ will be used to compute the loop
exponent.

The given $Z$-factors now allow to determine the flow of $u_{\psi}$
and $u_{\varphi\psi}$ caused by the diagrams B and C,
\begin{align}
\beta\left(u^{\mathrm{R}}_{\psi}\right) & =u^{\mathrm{R}}_{\psi}\left(-\epsilon+8u^{\mathrm{R}}_{\psi}+\tfrac{1}{2}\lambda^{2}_{\mathrm{R}}\right),\label{eq:beta_u_psi}\\
\beta\left(u^{\mathrm{R}}_{\varphi\psi}\right) & =u^{\mathrm{R}}_{\varphi\psi}\left(-\epsilon+4u^{\mathrm{R}}_{\varphi\psi}+2u^{\mathrm{R}}_{\varphi}+2u^{\mathrm{R}}_{\psi}+\tfrac{1}{4}\lambda^{2}_{\mathrm{R}}\right).\label{eq:beta_u_phi_psi}
\end{align}
The expressions agree with $\beta\left(u_{\varphi}\right)$ for $u_{\varphi}=u_{\psi}=u_{\varphi\psi}$,
only the $\lambda^{2}_{\mathrm{R}}$ due to $Z_{\psi}$ is new.

\begin{figure}
\centering
\includegraphics[scale=0.4]{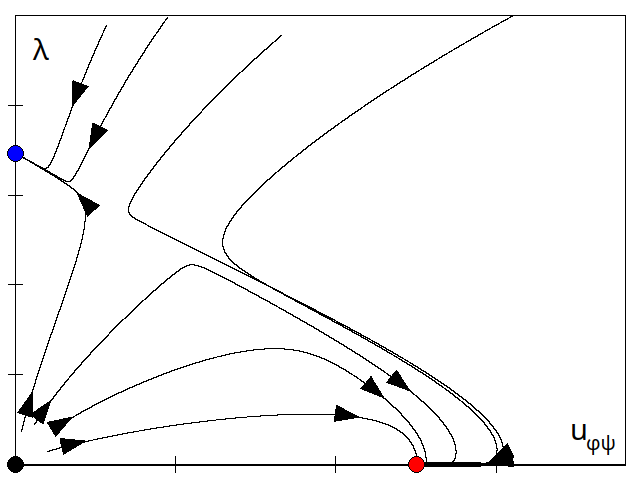}\caption{\label{fig:Flow}The flow of $u_{\psi}$, $u_{\varphi\psi}$ and $\lambda$
projected into the $u_{\varphi\psi}$-$\lambda$-plane. The initial
value of $u_{\psi}$ is $1/8$. There are two stable fixed points:
the usual excluded-volume fixed point with $\lambda=0$ and a PS-specific
fixed point with $\lambda\protect\neq0$ and $u_{\varphi}\protect\neq0$.}
\end{figure}

The most complicated part is the beta-function for the $\psi\varphi^{2}$
interactions generated by the PS diagram of Fig.~\ref{fig:Scheraga_Gamma}
and diagrams D and E of Fig.~\ref{fig:OneLoopExcl},
\begin{equation}
\beta\left(\lambda_{\mathrm{R}}\right)=\tfrac{\lambda_{\mathrm{R}}}{2}\left(-\epsilon+\tfrac{1}{4}\lambda^{2}_{\mathrm{R}}+2u^{\mathrm{R}}_{\varphi}+8u^{\mathrm{R}}_{\varphi\psi}\right).\label{eq:beta_lambda}
\end{equation}
The beta-functions (\ref{eq:beta_u_phi}, \ref{eq:beta_u_psi}, \ref{eq:beta_u_phi_psi},
\ref{eq:beta_lambda}) define a flow in a four-dimensional parameter
space. Since $\beta\left(u_{\varphi}\right)$ does not depend on other
parameters one can insert the stable fixed point value $u^{\mathrm{R}}_{\varphi}=\epsilon/8$
into the other equations. The projection of the remaining flow for
a typical $u_{\psi}$ into the $u_{\varphi\psi}$-$\lambda$ plane
is shown in Fig.~\ref{fig:Flow}.

Depending on the initial conditions one of two stable fixed points
is reached. According to the flow equations a strong repulsion (large
$u$) counteracts the pairing and leads to the conventional de Gennes
fixed point 
\begin{equation}
\lambda_{\mathrm{R}}=0,\qquad u^{\mathrm{R}}_{\varphi}=u^{\mathrm{R}}_{\psi}=u^{\mathrm{R}}_{\varphi\psi}=\epsilon/8.\label{eq:FpDeGennes}
\end{equation}
Because of $\lambda_{\mathrm{R}}=0$ this fixed point cannot describe
the denaturation transition, indicating that the transition is first
order.

When the repulsion is weak or $\lambda^{2}$ is large then the fixed
point 
\begin{align}
\lambda^{2}_{R}=3\epsilon,\qquad u^{\mathrm{R}}_{\varphi} & =\epsilon/8,\:u^{\mathrm{R}}_{\psi}=u^{\mathrm{R}}_{\varphi\psi}=0\label{eq:NewFixedPoint}
\end{align}
is reached. At this fixed point only the self-repulsion $u_{\varphi}$
of the single strand polymer is relevant. These results are valid
in linear order in $\epsilon$. A two-loop computation would provide
a more reliable extrapolation to $d=3$.

\subsection{Loop exponent}

The loop exponent can be deduced from the solution of the RG differential
equation~\cite{Amit1978} for the vertex function $\Gamma^{R}_{\tilde{\psi}\psi}$,
\begin{align}
\Gamma^{R}_{\tilde{\psi}\psi}\left(\boldsymbol{k},\omega\right) & =Z^{-2}_{\psi}\left(\mu\right)Z^{-1}_{s}\left(\mu\right)\mu^{2}F\left(\boldsymbol{k}/\mu,Z_{s}\left(\mu\right)\omega/\mu^{2},u_{R}\left(\mu\right),\lambda_{R}\left(\mu\right),W_{R}\left(\mu\right)\right)\label{eq:scalingFunc}\\
 & \sim\mu^{1/\nu-2\eta_{\psi}}F\left(\boldsymbol{k}/\mu,X\omega/\mu^{1/\nu},u_{R}\left(\mu\right),\lambda_{R}\left(\mu\right),W_{R}\left(\mu\right)\right).\nonumber 
\end{align}
All arguments of the scaling function $F$ are dimensionless, and
$u=\left\{ u_{\varphi},u_{\psi},u_{\varphi\psi}\right\} $. In the
second line we have inserted the expressions for the $Z$ factors,
and $X$ is a non-universal constant. For $\boldsymbol{k}=0$ and
small $\mu=\left(X\omega\right)^{\nu}$ it follows
\[
\Gamma_{\tilde{\psi}\psi}\left(0,\omega\right)\propto\omega^{1-2\eta_{\psi}\nu}F\left(0,1,u_{R},\lambda_{R},W_{R}\right).
\]
The Fourier transformation from $\omega$ to $s$ leads to
\begin{equation}
\Gamma_{\tilde{\psi}\psi}\left(\boldsymbol{k}=0,s\right)\sim s^{-\left(2-2\eta_{\psi}\nu\right)}.\label{eq:c_generic}
\end{equation}
The loop exponent at the fixed point (\ref{eq:NewFixedPoint}) with
finite $\lambda$ thus is 
\begin{equation}
c=2-2\eta_{\psi}\nu\cong2-\epsilon/4<2.\label{eq:c_new}
\end{equation}
This is in accord with a continuous phase transition of the type shown
in Fig.~\ref{fig:m_phi}. 

Near the de Gennes fixed point (\ref{eq:FpDeGennes}) the scaling
function $F$ contains the factor $\lambda^{2}_{R}\left(\mu\right)$.
The flow equation (\ref{eq:beta_lambda}) gives $\lambda_{R}\left(\mu\right)\sim\mu^{\omega_{\lambda}}$
with the crossover exponent $\omega_{\lambda}\cong$$\epsilon/8$.
The loop exponent follows as
\begin{equation}
c=2+2\left(\omega_{\lambda}-\eta_{\psi}\right)\nu\cong2+\epsilon/4>2.\label{eq:c_Gennes}
\end{equation}
This confirms that the denaturation is first order. The exponents
(\ref{eq:c_new}) and (\ref{eq:c_Gennes}) do not quite agree with
that of \cite{Pelity_2000}, derived with other scaling arguments.

\subsection{Other universality classes}

Other gPS variants allow some pairing of nucleotides with not exactly
matching length indices. The pairing interactions are then no longer
local in length space, and the double strand field has two length
variables. An extreme case is two complementary RNA molecules with
a periodic base sequence like $\mathrm{AGAG}\ldots$, which can pair
with arbitrary odd length offsets. This leads to the universality
class of conventional branched polymers~\cite{Lubensky1979,Dengler2024}
with upper critical dimension $8$. The excluded-volume interaction
is irrelevant in this case. The critical exponent $\nu$ is exactly
known~\cite{Parisi1981} for $d=3$ to be $\nu=1/2$.

\section{Conclusion}

The field theory describes the long-distance physics of the PS model
universality class, with or without excluded-volume. Without excluded-volume,
the PS denaturation loop reduces to a one-loop diagram, leading to
an exact closed-form solution for the phase diagram in three dimensions. 

With excluded-volume, one must distinguish the coupling constants
for self- and mutual exclusion, which contribute asymmetrically. The
unrenormalized constants for single and double strand RNA in general
are different.

Usually the one-loop RG calculation at least qualitatively also describes
the physics in three dimensions. In principle, however, it cannot
be ruled out that the flow diagram changes qualitatively between $d=4$
and $d=3$. A two-loop computation would provide a more accurate picture.

\bigskip{}

\bibliographystyle{habbrv}
\bibliography{DoublePolymer_Synopsis}

@Article{Poland_1966,
  author  = {D. Poland and H. A. Scheraga},
  title   = {Phase transitions in one dimension and the helix-coil transition in polyamino acids},
  doi     = {10.1021/j150451a002},
  journal = {J. Chem. Phys.},
  year    = {1966},
  pages   = {1456 - 1463},
  volume  = {45},
}

@Article{Gennes1972,
  author  = {P. G. de Gennes},
  title   = {Exponents for the excluded volume problem as derived by the {Wilson} method},
  doi     = {10.1016/0375-9601(72)90149-1},
  journal = {Phys. Lett. A},
  year    = {1972},
  number  = {5},
  pages   = {339 - 340},
  volume  = {38},
}

@Book{Amit1978,
  author    = {D. J. Amit and V. Martin-Mayor},
  title     = {Field theory, the renormalization group, and critical phenomena},
  doi       = {10.1142/5715},
  publisher = {World Scientific},
  address   = {},
  year      = {2005},
  issn      = {},
  editor    = {},
}

@Article{HH77,
  author  = {P. C. Hohenberg and B. I. Halperin},
  title   = {Theory of dynamic critical phenomena},
  doi     = {10.1103/RevModPhys.49.435},
  journal = {Rev. Mod. Phys.},
  year    = {1977},
  number  = {3},
  volume  = {49},
  pages   = {435-479},
}

@Article{Lubensky1979,
  author  = {T. C. Lubensky and J. Isaacson},
  title   = {Statistics of lattice animals and dilute branched polymers},
  doi     = {10.1103/PhysRevA.20.2130},
  journal = {Phys. Rev. A},
  year    = {1979},
  pages   = {2130 - 2146},
  volume  = {20},
  issue   = {5},
}

@Article{Parisi1981,
  author  = {G. Parisi and N. Sourlas},
  title   = {Critical Behavior of Branched Polymers and the {Lee-Yang} Edge Singularity},
  doi     = {10.1103/PhysRevLett.46.871},
  journal = {Phys. Rev. Lett.},
  year    = {1981},
  pages   = {871 - 874},
  volume  = {46},
  issue   = {14},
}

@Article{Pelity_1985,
  author  = {L. Peliti},
  title   = {Path integral approach to birth-death processes on a lattice},
  doi     = {10.1051/jphys:019850046090146900},
  journal = {J. Phys. France},
  year    = {1985},
  number  = {46},
  pages   = {1469-1483},
}

@Article{Pelity_2000,
  author  = {Y. Kafri and D. Mukamel and L. Peliti},
  title   = {Why is the {DNA} denaturation transition first order?},
  doi     = {10.1103/PhysRevLett.85.4988},
  journal = {Phys. Rev. Lett.},
  year    = {2000},
  number  = {23},
  volume  = {85},
  pages   = {4988-4991},
}

@Article{Garel_2001,
  author  = {T. Garel and C. Monthus and H. Orland},
  title   = {A simple model for {DNA} denaturation},
  doi     = {10.1209/epl/i2001-00391-2},
  journal = {Europhys. Lett.},
  year    = {2001},
  number  = {55},
  volume  = {1},
  pages   = {132-138 },
}

@Article{Garel_2004,
  author  = {T. Garel and H. Orland},
  title   = {Generalized {Poland-Scheraga} model for DNA hybridization. Biopolymers},
  doi     = {10.1002/bip.20140},
  journal = {Biopolymers},
  year    = {2004},
  number  = {75},
  volume  = {},
  pages   = {453-467},
  todo = {DOI!},
}

@Article{Richard_2004,
  author  = {C. Richard and A. J. Guttmann},
  title   = {{Poland-Scheraga} models and the {DNA} denaturation transition},
  doi     = {10.1023/B:JOSS.0000022370.48118.8b},
  journal = {J. Stat. Phys.},
  year    = {2004},
  pages   = {1456 - 1463},
  volume  = {115},
  number  = {3/4},
}

@Article{Giaco_2017,
  author  = {G. Giacomin  and M. Khatib},
  title   = {Generalized {Poland-Scheraga} denaturation model and two-dimensional renewal processes},
  doi     = {10.1016/j.spa.2016.06.017},
  journal = {Stoch. Proc. Appl.},
  year    = {2017},
  pages   = {526-573},
  volume  = {127},
  number  = {},
}

@Article{Berger_2024,
  author  = {Q. Berger and A. Legrand},
  title   = {Scaling limit of the disordered generalized {Poland-Scheraga} model for {DNA} denaturation},
  doi     = {10.1007/s00440-024-01304-1},
  journal = {Probab. Theory Relat. Fields},
  year    = {2024},
  pages   = {179-258},
  volume  = {190},
  number  = {},
  toDo = {},
}

@Article{Dengler2024,
  author  = {R. Dengler},
  title   = {Universality class of interacting directed single- and double-strand homopolymers},
  doi     = {10.1140/epje/s10189-024-00461-4},
  journal = {Eur. Phys. J. E},
  year    = {2024},
  volume  = {47},
  numer  =  {},
  pages   = {66},
}

@Article{Dengler2025,
  author  = {R. Dengler},
  title   = {Pseudo-{RNA} with parallel aligned single-strands and periodic base sequence as a new universality class},
  doi     = {10.1007/s10955-025-03477-y},
  journal = {J. Stat. Phys},
  year    = {2025},
  volume  = {192},
  numer  =  {},
  pages   = {95},
}
\bigskip{}

\appendix

\section{Appendix}

\subsection{Field theory on a lattice}

\label{subsec:FT_Lattice}The action (\ref{eq:Act}) can be derived
in a formal way by using nilpotent operators creating and annihilating
polymer ends on a lattice \cite{Dengler2025}. This method also reproduces
the excluded-volume interactions. A more direct way to understand
the action (\ref{eq:Act}) is to start with a version discretized
on a lattice. It suffices for this purpose to restrict the considerations
to the $\psi$-polymer alone,
\begin{align*}
Z_{(\psi)} & =\int^{i\infty}_{-i\infty}\mathrm{D}\tilde{\psi}\int^{\infty}_{-\infty}\mathrm{D}\psi e^{-S_{\psi}},\\
S_{(\psi)} & =\sum_{\mu}\left[\sum_{ij}\tilde{\psi_{i}}^{\mu}e^{-\beta E}v^{-1}_{ij}\psi^{\mu+1}_{j}-r_{2}\sum_{i}\tilde{\psi}^{\mu}_{i}\psi^{\mu}_{i}\right]+O\left(\psi^{4}\right).
\end{align*}
The fields $\psi$ have a lattice index $i\in\mathbb{Z}^{d}$ and
a length index $\mu\in\mathbb{Z}$. The quantity $v_{ij}$ is the
next-neighbor matrix. The $v^{-1}$ reproduces the Laplace operator
and a constant in the continuum limit.

Of primary interest is the bilinear part of $S$. The inverse of the
matrix defining the bilinear term is the propagator $G$. It is easy
to verify that this propagator counts paths on the lattice. With the
help of a matrix $A^{\ell}_{\mu\nu}=\delta_{\mu+\ell,\nu}$ it follows
by expanding in $r_{2}$
\[
\boldsymbol{G}_{\mu\nu}=\left(\boldsymbol{A}^{1}e^{-\beta E}\boldsymbol{v}^{-1}-r_{2}\boldsymbol{1}\right)^{-1}_{\mu\nu}=\sum^{\infty}_{\ell=1}\boldsymbol{v}^{\ell}e^{\beta E\ell}r^{\ell-1}_{2}\delta_{\mu-\ell,\nu}.
\]
The propagator allows to connect two arbitrary lattice points $i,j$
with length indices $\mu$ and $\nu$ differing by a length $\ell$.

The power of the next neighbor matrix $\boldsymbol{v}$ approaches
a Gaussian distribution with standard deviation $\sqrt{\ell/d}$ when
$\ell$ is large. The continuum limit propagator for a given distance
$\ell$ thus is a Gaussian function in space with a weight proportional
to $\ell^{-d/2}e^{-r_{0}\ell-\boldsymbol{x}^{2}/2\ell}$, where $r_{0}=c_{0}-\beta E.$
The $\psi\varphi^{2}$ PS interactions are additional information,
which must be added manually at the level of the lattice model or
the field theory.

A remark is in order concerning the path integral for the action (\ref{eq:Act})
with $u>0$. The perturbative renormalization proceeds via an expansion
around $u=0$, which requires only that the derivatives $\partial^{n}/\partial u^{n}|_{u=0}$
exist. At second sight, however, a problem appears to be for $u_{\varphi}>0.$
The path integral 
\[
\int^{\infty}_{-\infty}\mathrm{D}\varphi\int^{i\infty}_{-i\infty}\mathrm{D}\tilde{\varphi}\exp\left(-u_{\varphi}\ldots+\ldots\right)
\]
diverges, since for imaginary $\tilde{\varphi}$ the term $-u_{\varphi}\left(\tilde{\varphi}\varphi\right)^{2}$
is positive. This apparent difficulty arises because in (\ref{eq:Act})
we have retained only the leading nonlinearities of the excluded-volume
interaction
\[
S_{1}=-\int_{x}\ln\left(1+\int_{s}\tilde{\varphi}\varphi+\int_{s}\tilde{\psi}\psi\right),
\]
as generated by the lattice model~\cite{Dengler2025}. The full,
exact $S_{1}$, is only logarithmic in the fields and the path integral
converges; consequently, the perturbation series also exists for $u_{\varphi}>0$.
A dimensional analysis of this series near the critical point then
shows that the higher nonlinearities are irrelevant in the RG sense.
This justifies omitting from the outset terms such as $\int_{x}\left(\int_{s}\tilde{\varphi}\varphi\right)^{3}$
in (\ref{eq:Act}) in the perturbative renormalization. Such terms
would be essential in a functional RG approach. In short, the formally
correct order is: convergent path integral, perturbation series and
dimensional analysis. Irrelevant nonlinear terms can then be dropped
without any formal difficulty.

\subsection{Fourier transformation to length space}

\label{subsec:FT_integral}In three dimensions the $\psi$ propagator
\begin{equation}
G_{3\mathrm{d}}\left(s\right)=\int^{\infty}_{-\infty}\tfrac{\mathrm{d}\omega}{2\pi}e^{-i\omega s}\frac{1}{T-i\omega+L\sqrt{2r_{0}-i\omega}}\label{eq:FreqInt}
\end{equation}
from Eq.~(\ref{eq:G_3d_s}) can be Fourier transformed to length
space in closed form. The first step is to simplify the numerator
$N$. Defining $y^{2}=2r_{0}-i\omega$ it follows
\begin{align}
N & =y^{2}+T-2r_{0}+Ly=\left(y-y_{1}\right)\left(y-y_{2}\right),\label{eq:Numerat}\\
y_{1,2} & =\tfrac{1}{2}\left(-L\pm\sqrt{L^{2}+8r_{0}-4T}\right),\nonumber \\
\tfrac{1}{N} & =\tfrac{1}{y_{1}-y_{2}}\left(\tfrac{1}{\sqrt{2r_{0}-i\omega}-y_{1}}-\tfrac{1}{\sqrt{2r_{0}-i\omega}-y_{2}}\right).\nonumber 
\end{align}
The last line is the decomposition of $1/N$ into partial fractions.
 It remains to calculate integrals of the type 
\begin{align*}
J & =\int^{\infty}_{-\infty}\tfrac{\mathrm{d}\omega}{2\pi}\frac{e^{-i\omega s}}{\sqrt{B-i\omega}-y}=\int^{\infty}_{-\infty}\tfrac{\mathrm{d}\omega}{2\pi}e^{-i\omega s}\sum^{\infty}_{m=0}y^{m}\left(B-i\omega\right)^{-\tfrac{m+1}{2}}=\sum\tfrac{e^{-Bs}}{\Gamma\left(\left(m+1\right)/2\right)}y^{m}s^{\tfrac{m+1}{2}-1}\\
 & =e^{-Bs}s^{-1/2}\sum^{\infty}_{m=0}\tfrac{\left(y\sqrt{s}\right)^{m}}{\Gamma\left(\left(m+1\right)/2\right)}=e^{-Bs}\left[\frac{1}{\sqrt{\pi s}}+ye^{y^{2}s}\mathrm{erfc}\left(-y\sqrt{s}\right)\right].
\end{align*}
In the final step it was used $\sum^{\infty}_{m=0}\tfrac{x^{m}}{\Gamma\left(\left(m+1\right)/2\right)}=\frac{1}{\sqrt{\pi}}+xe^{x^{2}}\mathrm{erfc}\left(-x\right)$.
Putting everything together
\begin{align*}
G_{3\mathrm{d}}\left(s,r_{0},T\right) & =\theta\left(s\right)\tfrac{e^{-2r_{0}s}}{y_{1}-y_{2}}\left(y_{1}e^{y^{2}_{1}s}\mathrm{erfc}\left(-y_{1}\sqrt{s}\right)-y_{2}e^{y^{2}_{2}s}\mathrm{erfc}\left(-y_{2}\sqrt{s}\right)\right).
\end{align*}
The integral (\ref{eq:FreqInt}) originally runs along a path above
the branch cut at $\omega=-2ir_{0}$ and the pole. Shifting the integration
path upwards according to $\omega\rightarrow\omega+2iR$ gives the
functional equation
\[
G\left(s,r_{0},T\right)=e^{2Rs}G\left(s,r_{0}+R,T+2R\right).
\]
For a numeric evaluation of (\ref{eq:propPsi}) for generic $d$ one
could close the integration path in the lower complex $\omega$ half-plane.
There remains a contribution from a pole and a rapidly convergent
integral along the branch cut.

\end{document}